\documentclass[11pt]{article}
\usepackage{a4wide}
\makeatletter

\newcommand{\LyX}{L\kern-.1667em\lower.25em\hbox{Y}\kern-.125emX\@}

\newcommand{\lyxaddress}[1]{
  \par {\raggedright #1 
  \vspace{1.4em}
  \noindent\par}
}

\makeatother

\begin{document}

\title{\textbf{\large Qubit and Entanglement assisted Optimal Entanglement Concentration }\large }

\author{{\normalsize Somshubhro Bandyopadhyay}\thanks{
dhom@bosemain.boseinst.ernet.in
} {\normalsize }\normalsize }

\maketitle

\lyxaddress{\hfill{}Department of Physics, Bose Institute, 93/1 A. P. C Road, Calcutta
- 700009, India\hfill{} }

\begin{abstract}
We present two methods for optimal entanglement concentration from pure entangled
states by local actions only. However a prior knowledge of the Schmidt coefficients
is required. The first method is optimally efficient only when a finite ensemble
of pure entangled states are available whereas the second method realizes the
single pair optimal concentration probability. We also propose an entanglement
assisted method which is again optimally efficient even for a single pair. We
also discuss concentrating entanglement from N-partite cat like states. 
\end{abstract}

\section{Introduction}

Quantum superposition principle gives rise to what is known as quantum entanglement
\cite{1}, a non classical property exhibited by composite systems. By virtue
of this property, subsystems of a composite system show nonlocal correlations
between them and had been studied extensively in the context of EPR problem
\cite{2}and Bell's inequality \cite{3}. However rapid developments in the
last few years changed the scenario altogether. Now it is well understood that
entanglement serves as an useful physical resource for information processing
\cite{4}, and quantum computation \cite{5} and allows manipulation like any
other physical resources. Some key applications of entanglement include, quantum
teleportation \cite{6}, dense coding \cite{7}, secure key distribution \cite{8}
and reduction of communication complexity \cite{9}. Here one may note that
maximally entangled states (Bell states) are essential for faithful quantum
communication, for example teleportation \cite{6} and secure quantum key distribution
\cite{8}. Therefore, protocols have been developed for obtaining a better entangled
state from a less entangled one by local operations and classical communications.
These processes are suitably termed as entanglement concentration \cite{10,11,12,13}
when one extracts maximally entangled states (henceforth MES) from pure entangled
states and purification or distillation \cite{11,14,15} when MES are obtained
from mixed entangled states. 

The basic idea of entanglement concentration is the following: Two distant observers,
Alice and Bob, are supplied with a finite ensemble of pure states from which
they wish to extract maximum possible MES, where they are only allowed to perform
local actions, e.g., unitary transformations and measurements, on their respective
subsystems along with any auxiliary system (ancilla) they might prepare and
classical communication. 

The aim of this contribution is to present two methods for optimal entanglement
concentration from pure entangled states using only local actions. For our methods
to be successful Alice and Bob should know the Schmidt coefficients of the given
entangled state(s). The first method becomes optimally efficient when a finite
(not necessarily large) ensemble of pure states are available. The second method
that we suggest however produces the optimal single pair concentration probability
and possibly powerful than the first method. Besides we also propose an entanglement
assisted concentration protocol. We show that if Alice uses an entangled state
as an ancillary resource (for example, as in the protocol of Bose et. al. \cite{12})
then one can obtain the optimal single pair concentration probability by opting
for a different measurement scheme. We also discuss how the methods developed
for entanglement concentration for bipartite systems can also be successfully
applied in case of multipartite cat-like states. 

\emph{Qubit assisted methods:} The first method that we suggest requires Alice
to prepare a qubit (ancilla: an auxiliary two level quantum system) in a state,
say, \( \left| \chi \right\rangle  \) (the coefficients of this state are initially
chosen to be the Schmidt coefficients of the supplied entangled state). The
procedure needs to be carried out separately on each member of the given ensemble.
Thus, the ancilla qubit after being used once to purify a single pair, is brought
back to the desired state by passing it through a polarizer for further application.
Here we would like to point out that in order to obtain the optimal fraction
of MES the method should be continued in an iterative fashion, in principle,
\textit{indefinitely}. Let us explain what we mean by this. Suppose Alice and
Bob are initially supplied with \( N \) (as we shall see need not be necessarily
very large) pure entangled states. After carrying out the protocol over all
the members of this ensemble they are left with say \( N_{1} \) number of MES
and \( (N-N_{1}) \) of less entangled pairs whereby they select the members
of this less entangled sub ensemble, repeat the protocol and so on. This iterative
process if continued indefinitely, Alice and Bob finally end up with the optimal
fraction of MES. It may be worth mentioning that the present method doesn't
require the supplied ensemble to be infinite (i.e., the optimal fraction is
not approached asymptotically), but in practice the iterative procedure makes
sense only when the supplied ensemble is reasonably large. Our method can also
be understood intuitively from conservation of entanglement. As will be shown
later that at every step of this concentration procedure average entanglement
remains conserved implying that as MES are being produced the remaining pairs
turn less entangled. Finally when the optimal fraction of the Bell states is
obtained in the limit of an infinite sequence, the remaining pairs become totally
disentangled. 

The second protocol goes like this: Let us assume that Bob takes the responsibility
of performing the desired local operations for entanglement concentration. He
now prepares an ancillary qubit in state \( \left| 0\right\rangle  \). The
procedure now works in two steps. The first step involves in performing a CNOT
on the two qubits that Bob holds. The second step is to perform an optimal state
discrimination measurement (an optimal POVM) on any one of the qubits belonging
to Bob. Consequently a conclusive result of such a measurements generates a
maximally entangled state between Alice and Bob. 

\emph{Entanglement assisted method:} Here any one of the parties, say Alice
requires to prepare an entangled state to implement the protocol. In Ref. \cite{12}
the authors proposed an optimally efficient entanglement assisted concentration
protocol using entanglement swapping \cite{16}. However the method \cite{12}
is not optimally efficient for concentrating entanglement from a single pair.
We show that resorting to a different measurement scheme one can however obtain
the optimal single pair concentration probability. 

\emph{Multipartite entanglement Concentration:} Bipartite pure entangled states
have unique representation through their Schmidt decomposable property. This
makes dealing with bipartite pure states relatively easier than multipartite
states not because of the larger number of parties being involved in the later
case but for the fact that there is no unique representation for pure multipartite
entangled states analogous to Schmidt decomposition. In this paper we treat
the problem of multipartite entanglement concentration only for a restricted
class of states, viz. the N-partite cat-like states and one should note that
these type of states are Schmidt decomposable. One advantage of the methods
that we developed for treating bipartite systems is that they are equally applicable
for multipartite systems without any modifications whatsoever. Using them we
show that the probability of entanglement concentration for multipartite cat
like states is same as that in bipartite systems. Thus the obtained concentration
probability is conjectured to be optimal for multipartite systems that are Schmidt
decomposable. 

\emph{Tools required for entanglement concentration:- Local Operations and Classical
communication:} The local operations that are in general used for entanglement
concentration and distillation procedures include projective Von Neumann measurement,
generalized measurements, in particular the POVM required for optimal state
discrimination between two non orthogonal states \cite{17}, incomplete Bell
measurements (for example, see Ref. \cite{12}) and the CNOT (or quantum XOR)
gate (an unitary transformation acting on pairs of spin-1/2 that flips the second
spin if and only if the first spin is ``up'' i.e., it changes the second bit
Iff the first bit is ``1'' \footnote{
In our notation \( \left| \uparrow \right\rangle =\left| 1\right\rangle  \)
and \( \left| \downarrow \right\rangle =\left| 0\right\rangle  \).
} and is defined by the following transformation rules: \( \left| 00\right\rangle \rightarrow \left| 00\right\rangle ;\left| 01\right\rangle \rightarrow \left| 01\right\rangle ;\left| 10\right\rangle \rightarrow \left| 11\right\rangle ;\left| 11\right\rangle \rightarrow \left| 10\right\rangle  \)).

Besides these, classical communication is an integral part of all protocols.
It can be either two way or one way depending on the respective protocol. This
is necessary to inform the partners about the result of the local quantum operations
in order to select the successful cases. 

This paper is arranged as follows. In Sec. 2 we present the qubit assisted entanglement
concentration methods. In Sec. 3 we discuss entanglement assisted entanglement
concentration. We propose a measurement scheme that produces the optimal single
pair concentration probability. Sec. 4 is devoted to discussions regarding the
relative merits of our schemes compared to the existing protocols \cite{10,11,12}.
Experimental feasibility of the suggested and the existing methods is also discussed.
In Sec. 5 entanglement concentration from multipartite cat-like states is discussed.
Finally in Sec. 6 we summarize and conclude.

\section{Qubit assisted Entanglement Concentration}

\subsection{Proposal one\emph{: }}

Suppose Alice and Bob share a pure entangled state of the form,

\begin{equation}
\label{1}
\left| \Phi \right\rangle _{AB}=\alpha \left| 00\right\rangle _{AB}+\beta \left| 11\right\rangle _{AB}
\end{equation}

where we take \( \alpha ,\beta  \) to be real and \( \alpha <\beta  \). 

Alice prepares a qubit in the state,
\begin{equation}
\label{2}
\left| \chi \right\rangle _{A}=\alpha \left| 0\right\rangle +\beta \left| 1\right\rangle .
\end{equation}

The preparation of the qubit in state (2) is crucial. Note that Alice should
know the Schmidt coefficients of the supplied pure entangled state in order
to prepare her ancillary qubit. Thus the combined state of the three qubits
is given by,

\begin{equation}
\label{3}
\left| \Psi \right\rangle _{AB}=\left| \chi \right\rangle _{A}\otimes \left| \Phi \right\rangle _{AB}=\alpha ^{2}\left| 000\right\rangle _{A_{1}A_{2}B}+\alpha \beta \left| 011\right\rangle _{A_{1}A_{2}B}+\alpha \beta \left| 100\right\rangle _{A_{1}A_{2}B}+\beta ^{2}\left| 111\right\rangle _{A_{1}A_{2}B}
\end{equation}

The first two qubits belongs to Alice (denoted by \( A_{1} \) and \( A_{2} \))
and the last one belongs to Bob. The entanglement concentration procedure involves
two steps.

\textit{Step 1}: Alice performs a CNOT operation on her two qubits. Bob doesn't
need to do anything. This is the most difficult stage because to carry out CNOT
operation is in no sense a trivial job. The resulting state turns out to be

\begin{equation}
\label{4}
\left| \Psi ^{\prime }\right\rangle _{AB}=\alpha ^{2}\left| 000\right\rangle _{A_{1}A_{2}B}+\alpha \beta \left| 011\right\rangle _{A_{1}A_{2}B}+\alpha \beta \left| 110\right\rangle _{A_{1}A_{2}B}+\beta ^{2}\left| 101\right\rangle _{A_{1}A_{2}B}
\end{equation}

Interchanging the position of the first two qubits since both belong to Alice
Eq. (4) can be written as, 
\begin{equation}
\label{5}
\left| \Psi ^{\prime }\right\rangle _{AB}=\alpha ^{2}\left| 000\right\rangle _{A_{2}A_{1}B}+\alpha \beta \left| 101\right\rangle _{A_{2}A_{1}B}+\alpha \beta \left| 110\right\rangle _{A_{2}A_{1}B}+\beta ^{2}\left| 011\right\rangle _{A_{2}A_{1}B}
\end{equation}

\textit{Step 2}: This is an easy part where Alice performs a Von Neumann projective
measurement on the qubit \( A_{2} \), she holds i.e., she measures the z-component
of the spin of qubit \( A_{2} \). This is brought about by writing Eq. (5)
as,

\begin{equation}
\label{6}
\left| \Psi ^{'}\right\rangle _{AB}=\left| 0\right\rangle _{A_{2}}\otimes [\alpha ^{2}\left| 00\right\rangle +\beta ^{2}\left| 11\right\rangle ]_{A_{1}B}+\alpha \beta \left| 1\right\rangle _{A_{2}}\otimes [\left| 01\right\rangle +\left| 10\right\rangle ]_{A_{1}B}
\end{equation}

Thus if the outcome of Alice's measurement is ``up '' i.e ``1'', the resulting
pair shared by Alice and Bob gets maximally entangled. Otherwise they come up
with a lesser entangled state than what they initially shared. So the question
is performing the above operations how often they succeed in getting a maximally
entangled state. This can easily be seen by noting that the probability with
which outcome ``1'' is obtained is \( 2\alpha ^{2}\beta ^{2} \). This is
in fact the single pair concentration probability using this method. However
this is not the optimal probability. We now show that given a finite number
of entangled states one can implement an iterative procedure to obtain the optimal
fraction of maximally entangled states. 

Suppose Alice and Bob initially shared N (which we shall presently see need
not necessarily be very large) pure entangled states. The basic steps are the
following: 

(1) Applying the protocol over the \( N \) members individually, they end up
with \( 2N\alpha ^{2}\beta ^{2} \) number of MES. 

(2) Now they pick out the remaining \( N(1-2\alpha ^{2}\beta ^{2})=N(\alpha ^{4}+\beta ^{4}) \)
number of pairs which are not maximally entangled. Note that now each member
of this less entangled sub ensemble are in a state given by

\begin{equation}
\label{7}
\left| \Phi _{1}\right\rangle _{AB}=\alpha _{1}\left| 00\right\rangle _{AB}+\beta _{1}\left| 11\right\rangle _{AB},
\end{equation}

where \( \alpha _{1}=\frac{\alpha ^{2}}{\sqrt{\alpha ^{4}+\beta ^{4}}} \) and
\( \beta _{1}=\frac{\beta ^{2}}{\sqrt{\alpha ^{4}+\beta ^{4}}} \). Accordingly,
Alice prepares her qubit in the state 
\begin{equation}
\label{8}
\left| \chi _{1}\right\rangle _{A}=\alpha _{1}\left| 0\right\rangle +\beta _{1}\left| 1\right\rangle .
\end{equation}
and the single pair concentration procedure is applied again. 

(3) This iterative procedure is continued \textit{indefinitely}.

Now we show that the above procedure, when continued indefinitely, in the limit
of an infinite sequence, the final ensemble generated comprise \( 2\beta ^{2} \)
fraction of MES. 

The proof is as follows: If they begin with N pair of pure entangled states
and finally end up with \( N_{ME} \) number of MES, then the fraction of MES
produced is given by,

\begin{equation}
\label{9}
\frac{N_{ME}}{N}=\left[ 2\alpha ^{2}\beta ^{2}+\frac{2\alpha ^{4}\beta ^{4}}{\left( \alpha ^{4}+\beta ^{4}\right) }+\frac{2\alpha ^{8}\beta ^{8}}{\left( \alpha ^{4}+\beta ^{4}\right) \left( \alpha ^{8}+\beta ^{8}\right) }+\frac{2\alpha ^{16}\beta ^{16}}{\left( \alpha ^{4}+\beta ^{4}\right) \left( \alpha ^{8}+\beta ^{8}\right) \left( \alpha ^{16}+\beta ^{16}\right) }+...\right] 
\end{equation}

which can be rewritten as,

\begin{equation}
\label{10}
\frac{N_{ME}}{N}=\left[ 2\alpha ^{2}\beta ^{2}+2\beta ^{4}\{\frac{1}{(1+x^{4})}+\frac{x^{4}}{\left( 1+x^{4}\right) \left( 1+x^{8}\right) }+\frac{x^{12}}{\left( 1+x^{4}\right) \left( 1+x^{8}\right) \left( 1+x^{16}\right) }+...\}\right] 
\end{equation}

where \( 0<x=\frac{\beta }{\alpha }<1 \). 

It is straightforward to show that the following infinite series 

\begin{equation}
\label{11}
I=\frac{1}{(1+x^{4})}+\frac{x^{4}}{(1+x^{4})(1+x^{8})}+\frac{x^{12}}{(1+x^{4})(1+x^{8})(1+x^{16})}+\frac{x^{28}}{(1+x^{4})(1+x^{8})(1+x^{16})(1+x^{32})}+...
\end{equation}

uniformly converges to 1 for all \( x\in (0,1) \), whereby \( \frac{N_{ME}}{N}=2\beta ^{2} \),
known to be the optimal fraction of MES obtainable from pure entangled states.
Hence our protocol indeed succeeds in extracting the optimal fraction of Bell
states from an arbitrary number of pure entangled states. The efficiency of
this method though optimal crucially depends on the rate of convergence of the
series (11). However it is easy to see from (11) that the series converges very
rapidly. From a practical point of view the optimal fraction is therefore approached
very fast starting with a reasonable number of pure entangled states. 

We now discuss the operational meaning of our protocol. We have seen that the
optimal fraction is \textit{independent} of the size of the ensemble. By this
we mean that the optimal fraction of Bell states that can be obtained is not
reached asymptotically i.e.. it is not necessary to have an infinite ensemble.
However, to achieve the optimal result the iteration procedure needs to be continued,
in principle, \textit{indefinitely}. However the rapid convergence of the series
(11) ensures that , even in practice, to continue this iterative procedure in
order to approach the optimal fraction we only need to have a reasonably sized
ensemble. Note that for this method to be successful it is necessary to know
\( \alpha  \) and \( \beta  \), the Schmidt coefficients of the initially
supplied pure entangled states. Classical communication is also required for
Alice to convey her result to Bob in order to select the successful cases. 

Now we show that a particular measure of entanglement viz. entanglement of single
pair purification \cite{12}, is conserved on an average. We treat this conservation
of entanglement in the same sense as discussed in Ref. \cite{12}. We show that
in our case also average entanglement is indeed conserved and therefore optimal
in the sense that best combination of entangled states are obtained in the process.
From the results of Lo and Popescu \cite{13} it follows that initially the
average values of entanglement shared between Alice and Bob is 

\begin{equation}
\label{12}
\left\langle E\right\rangle _{before}=2\beta ^{2}
\end{equation}

where \( \beta  \) is the smaller Schmidt coefficient. After carrying out our
protocol on a single pair the average entanglement shared by Alice and Bob is
given by,

\begin{equation}
\label{13}
\left\langle E\right\rangle _{after}=2\beta ^{4}+2\alpha ^{2}\beta ^{2}=2\beta ^{2}
\end{equation}

Thus average entanglement is conserved at each step of the above procedure which
implies that when the optimal fraction is reached, the remaining fraction becomes
totally disentangled provided the process is continued indefinitely. 

Now a few remarks regarding the efficiency of our method as compared to the
other existing protocols \cite{10,11,12}. As we have discussed earlier, to
realize the optimal fraction of MES the iterative procedure needs to be continued
indefinitely. But \textit{in practice the iterative procedure makes sense only
when Alice and Bob have in their possession a reasonable number of pure entangled
states to start with}.  Therefore we can only say that our method is as efficient
as the other optimal ones \cite{10,11,12}. As noted earlier the optimal fraction
is approached very fast (see (11)) so any reasonably finite number of pure entangled
states is required to implement this method successfully. However we note that
a knowledge of Schmidt coefficients is necessary to implement our method and
Procrustean method \cite{10} whereas the Schmidt decomposition method, although
works for any unknown ensemble of pure states but there the optimal fraction
is approached asymptotically.

\subsection{Proposal two: }

Alice prepares an ancilla qubit in state \( \left| 0\right\rangle  \). Thus
the combined state is

\begin{equation}
\label{14}
\left| \Phi \right\rangle _{AB}\otimes \left| 0\right\rangle _{B}=\alpha \left| 000\right\rangle _{AB}+\beta \left| 110\right\rangle _{AB}
\end{equation}

where the first qubit belongs to Alice and the last two belongs to Bob. Bob
now subjects his two qubit to a CNOT operation whereby the new state given by

\begin{equation}
\label{15}
\left| \Psi \right\rangle _{AB}=\alpha \left| 000\right\rangle _{AB}+\beta \left| 111\right\rangle _{AB}
\end{equation}

can also be written as

\begin{equation}
\label{16}
\left| \Psi \right\rangle _{AB}=\frac{1}{\sqrt{2}}\left[ \left| \Phi ^{+}\right\rangle _{AB}\left( \alpha \left| 0\right\rangle +\beta \left| 1\right\rangle \right) _{B}+\left| \Phi ^{-}\right\rangle _{AB}\left( \alpha \left| 0\right\rangle -\beta \left| 1\right\rangle \right) _{B}\right] 
\end{equation}

where the states \( \left| \Phi ^{\pm }\right\rangle _{AB} \) are defined by,

\begin{equation}
\label{17}
\left| \Phi ^{\pm }\right\rangle _{AB}=\frac{1}{\sqrt{2}}(\left| 00\right\rangle _{AB}\pm \left| 11\right\rangle _{AB})
\end{equation}

From (16) it is clear that a state discrimination measurement which can conclusively
distinguish between the two non orthogonal states \( (\alpha \left| 0\right\rangle +\beta \left| 1\right\rangle ) \)
and \( (\alpha \left| 0\right\rangle -\beta \left| 1\right\rangle ) \) will
give the desired result. Now, this optimal state discrimination measurement
which is an optimal POVM measurement can be carried out on any one of the two
qubits that Bob holds and let us assume that it is the second qubit on which
such a measurement is performed. Note that the scalar product of these two nonorthogonal
states is \( \left( \alpha ^{2}-\beta ^{2}\right)  \)). The respective positive
operators that form an optimal POVM \cite{17} are:

\begin{equation}
\label{18}
A_{1}=\frac{1}{2\alpha ^{2}}\left( \begin{array}{cc}
\beta ^{2} & \alpha \beta \\
\alpha \beta  & \alpha ^{2}
\end{array}\right) ;A_{2=}\left( \begin{array}{cc}
\beta ^{2} & -\alpha \beta \\
-\alpha \beta  & \alpha ^{2}
\end{array}\right) ;A_{3}=\left( \begin{array}{cc}
1-\frac{\beta ^{2}}{\alpha ^{2}} & 0\\
0 & 0
\end{array}\right) 
\end{equation}

The optimal probability of obtaining a conclusive result from such a generalized
measurement (POVM) is \( 1-\left( \alpha ^{2}-\beta ^{2}\right) =2\beta ^{2} \).
It is clear that this is also being the probability of obtaining a maximally
entangled state shared by Alice and Bob because a conclusive outcome implies
that the entangled state shared by Alice and Bob is now given by either \( \left| \Phi ^{+}\right\rangle _{AB} \)
or \( \left| \Phi ^{-}\right\rangle _{AB} \) depending on the state of the
second qubit of Bob. For example, suppose Bob concludes that the state of his
second qubit after the POVM measurement is \( (\alpha \left| 0\right\rangle +\beta \left| 1\right\rangle ) \),
then with certainty he also concludes that the maximally entangled state that
he now shares with Alice is \( \left| \Phi ^{+}\right\rangle  \). Thus this
method produces the optimal probability of entanglement concentration for a
single pure entangled state.

\section{Entanglement assisted Entanglement Concentration: }

In the method that we now discuss Alice needs to prepare a similar entangled
state locally In Ref. \cite{12} Bose et. al. proposed an optimally efficient
protocol for entanglement concentration via entanglement swapping \cite{16}
where an ancillary entangled state is prepared beforehand to carry out the protocol.
 We note that the single pair concentration probability for a state of the form
(1) as discussed in Ref. \cite{12} is \( 2\alpha ^{2}\beta ^{2} \) and this
is not the optimal value. Here we would like to point out that the first proposal
of ours (see Sec. 2.1) succeeds in realizing the same single pair concentration
probability using only a single qubit as an additional resource.  Since an entanglement
is a more powerful resource than a qubit it is not unusual to suspect that a
better measurement scheme might be devised which can improve the single pair
concentration probability. This is what we suggest here. Of course the feasibility
to realize our method experimentally is not very certain taking into account
the present day technology. The advantage of the protocol of Bose et. al \cite{12}
is that their method can be successfully implemented in the laboratory with
the present day technology. 

We begin with the following facts. Alice prepares Alice and Bob share a pure
entangled state of the form (1). Alice also locally prepares another entangled
pair in the same state. Thus the combined state may be written as,

\begin{equation}
\label{19}
\left| \Psi \right\rangle _{AB}=\left| \Phi \right\rangle _{A}\otimes \left| \Phi \right\rangle _{AB}=\left( \alpha \left| 00\right\rangle _{A_{1}A_{2}}+\beta \left| 11\right\rangle _{A_{1}A_{2}}\right) \otimes \left( \alpha \left| 00\right\rangle _{A_{3}B}+\beta \left| 11\right\rangle _{A_{3}B}\right) 
\end{equation}
 where the suffices \( A_{1},A_{2} \) denote the qubits of the auxiliary entangled
pair and the suffix \( A_{3} \) denotes the qubit that is the part of the entangled
pair shared by Alice and Bob. Now we note that (19) can also be written as\\

\( \left| \Psi \right\rangle _{AB}=\frac{1}{\sqrt{2}}\left[ \left( \alpha ^{2}\left| 00\right\rangle +\beta ^{2}\left| 11\right\rangle \right) _{A_{2}A_{3}}\left| \Phi ^{+}\right\rangle _{A_{1}B}+\left( \alpha ^{2}\left| 00\right\rangle -\beta ^{2}\left| 11\right\rangle \right) _{A_{2}A_{3}}\left| \Phi ^{-}\right\rangle _{A_{1}B}\right]  \)

\begin{equation}
\label{20}
+\alpha \beta \left[ \left| \Psi ^{+}\right\rangle _{A_{2}A_{3}}\left| \Psi ^{+}\right\rangle _{A_{1}B}+\left| \Psi ^{-}\right\rangle _{A_{2}A_{3}}\left| \Psi ^{-}\right\rangle _{A_{1}B}\right] 
\end{equation}

Now the measurement part of Alice takes place in two steps:

Step 1: A measurement that projects the state onto either of the subspaces span
by \( \left\{ \left| 00\right\rangle ,\left| 11\right\rangle \right\}  \) or
\( \left\{ \left| 01\right\rangle ,\left| 10\right\rangle \right\}  \). 

Step 2: An appropriate measurement depending on the outcome of step 1 that generates
a maximally entangled state between Alice and Bob. 

First note that there are two possible outcomes of the measurement done in step
1 and consequently, measurement part of step 2 is to be defined accordingly. 

\emph{Outcome one}: Alice's measurement projects the state onto the subspace
spanned by \( \left\{ \left| 00\right\rangle ,\left| 11\right\rangle \right\}  \).
This happens with probability \( \alpha ^{4}+\beta ^{4} \). At this point Alice
needs to perform a state discrimination procedure to discriminate between the
two non orthogonal states (after normalization) \( \left( \left| \chi ^{+}\right\rangle _{A_{2}A_{3}}=\alpha _{1}\left| 00\right\rangle +\beta _{1}\left| 11\right\rangle \right)  \)
and \( \left( \left| \chi ^{-}\right\rangle _{A_{2}A_{3}}=\alpha _{1}\left| 00\right\rangle -\beta _{1}\left| 11\right\rangle \right)  \)
where \( \alpha _{1}=\frac{\alpha ^{2}}{\sqrt{\alpha ^{4}+\beta ^{4}}} \) and
\( \beta _{1}=\frac{\beta ^{2}}{\sqrt{\alpha ^{4}+\beta ^{4}}} \).  This generalized
measurements are performed on the qubits \( A_{2} \) and \( A_{3} \) jointly.
This is an important point to note. The optimal probability with which a conclusive
result is obtained by performing an optimal POVM measurement is 

\begin{equation}
\label{21}
p(conclusive)=1-\left| \left\langle \chi ^{+}\right| \left. \chi ^{-}\right\rangle \right| =\frac{2\beta ^{4}}{\alpha ^{4}+\beta ^{4}}
\end{equation}
It is clear from (20) that a conclusive result immediately implies that Alice
and Bob share a maximally entangled state. For example if Alice concludes with
certainty after the state discrimination measurement that the state is, say
\( \left| \chi ^{+}\right\rangle _{A_{2}A_{3}} \), then it immediately follows
that Alice and Bob now share the maximally entangled state \( \left| \Phi ^{+}\right\rangle _{A_{1}B} \).
We keep the suffices \( A_{i},B \) etc. in order to avoid any confusion. Note
that, given ``outcome 1'' has occurred the single pair purification probability
is just \( p(conclusive) \). 

\emph{Outcome two:} After performing the measurement defined in step 1  the
other possible outcome is: the state is projected onto the subspace spanned
by \( \left\{ \left| 01\right\rangle ,\left| 10\right\rangle \right\}  \).
This outcome occurs with probability \( 2\alpha ^{2}\beta ^{2} \). This result
when occurs actually simplifies the measurement part in step 2. Since now there
is now no need to perform a POVM measurement. The measurement that needs to
be performed in this case is an incomplete Bell measurement on the qubits \( A_{2} \)
and \( A_{3} \). It is clear from (22) that such a measurement always results
in a maximally entangled state (\( \left| \Psi ^{+}\right\rangle  \) or \( \left| \Psi ^{-}\right\rangle  \))
between Alice and Bob. Therefore, given ``outcome 2'' has occurred the single
pair purification probability is 1.

Now the question is: What is the efficiency of the above scheme ? Or, in other
words what is the single pair concentration probability ?

It is easy to obtain that the probability of single pair purification by implementing
the above method  denoted by \( p_{SPC} \) (SPC stands for single pair concentration)
is:

\begin{equation}
\label{22}
p_{SPC}=2\beta ^{4}+2\alpha ^{2}\beta ^{2}=2\beta ^{2}
\end{equation}

Thus the present method produces the optimal single pair concentration probability.
In this method the additional resource required is an entangled state. However
as we have seen in the previous section (Sec. 2.2) that to obtain the optimal
probability, an ancillary qubit is sufficient. This implies that the qubit assisted
method is a better one than the entanglement assisted method although both are
able to convert a pure entangled state to a MES optimally.

\section{Entanglement Concentration for N-partite Cat like states}

We now proceed to show how our scheme works for multipartite entangled states.
The method used above relied strongly on the existence of Schmidt decomposition
for bipartite states. The difficulty in treating multipartite entangled states
is that there are many possible forms of entanglement and there is no analogue
to the Schmidt decomposition of bipartite systems. We therefore deal in particular
with N-partite cat like pure entangled states. For simplicity let us first consider
the following three partite state,

\begin{equation}
\label{23}
\left| \Phi \right\rangle _{ABC}=\alpha \left| 000\right\rangle _{ABC}+\beta \left| 111\right\rangle _{ABC}
\end{equation}

Here our task becomes easier because the two proposals discussed in Sec. (2)
can also be successfully applied for concentrating entanglement from these multipartite
cat like states. Thus the methods for entanglement concentration from the state
(23) proceeds exactly the same way as discussed in Sec. 2.1 and Sec. 2.2. 

If we follow the scheme of Sec. 2.1 then Alice needs to prepare a qubit in the
state defined by Eq. (2). She then performs a CNOT operation on her two particles
and finally a Von Neumann projective measurement in the \{0,1\} basis. If the
result of her measurement is ``1'' which occurs with probability \( 2\alpha ^{2}\beta ^{2} \),
Alice, Bob and Carol then end up with a GHZ state of the form,

\begin{equation}
\label{24}
\left| \Phi \right\rangle ^{GHZ}_{ABC}=\frac{1}{\sqrt{2}}(\left| 011\right\rangle _{ABC}+\left| 100\right\rangle _{ABC})
\end{equation}

Thus it turns out given a finite ensemble of the three partite entangled states
of the form defined by (25) the maximum fraction of GHZ states obtainable is
\( 2\beta ^{2} \). 

We can also follow the method discussed in Sec. 2.2. and the result is the same.
The usefulness of the second proposal is that it does not require an ensemble
to become successful. Thus given a single multipartite entangled state of the
form (23), the probability with which one can successfully generate a GHZ state
is \( 2\beta ^{2} \). 

It is clear that our scheme is trivially generalized to purify N-partite states
of the form,

\begin{equation}
\label{25}
\left| \Phi \right\rangle _{1,2...N}=(\alpha \left| 00.....0\right\rangle _{1,2.....N}+\beta \left| 11....1\right\rangle _{1,2....N})
\end{equation}

where, the maximum fractional yield for a finite ensemble remain the same as
noted in case of bipartite systems. The probability that we obtain for concentrating
entanglement for N-partite cat like states having the form (25) is \( 2\beta ^{2} \)
and is conjectured to be optimal.

\section{Discussion}

Entanglement Concentrating procedures generate maximally entangled states which
can be used for quantum communication with highest efficiency. The protocols
that we discussed are state dependent in the sense that knowledge of the Schmidt
coefficients is required. 

It should be noted that the qubit assisted method discussed in Sec. 2.2 is better
than the entanglement assisted method although both the protocols are optimal
for a single pair. The advantage is two fold: First is it is easier to prepare
a qubit in any desired state (pass it through a Stern -Garlach apparatus appropriately
oriented) than to prepare an entangled state. The second advantage is more important.
The qubit can be reused once the operation is over for one pair. But in case
of entanglement assisted process the auxiliary entangled state needs to be prepared
for every individual pair because after a single operation the state gets destroyed. 

One important issue is how many pure states are available to carry out the concentration
protocols. It may so happen that only a limited number of entangled states are
available. In that case one has to resort to the single pair concentration protocols
and apply the methods on the members individually. However when an ensemble
of pure entangled states are available one may apply single pair protocols on
individual pairs or may use protocols that are not efficient for a single pair
but becomes optimally efficient for a large number of supplied states, for example,
the method suggested in Sec. 2.1. In this context an important issue is the
experimental feasibility of the protocols. 

For our methods to be successful we need a CNOT between one particle of the
entangled pair and ancilla. But this is not something that can be implemented
with photons as the technology stands today. On the other hand Procrustean method
\cite{10}, though it involves a POVM, can be implemented with a polarization
dependent beam splitter for photons. The scheme in Ref. \cite{12}, only needs
incomplete Bell state measurements. However for ions entangled in distant traps,
it is difficult to have a polarization dependent filter for the procrustean
method. To purify by entanglement swapping would mean involving two more trapped
ions. In such cases a scheme with only one ancilla ion on which only a CNOT
is to be made will be very helpful.

\section{Conclusions}

In fine we have described two optimal protocols for concentrating entanglement
from pure bipartite entangled states. The first method becomes optimally efficient
only when a reasonable number of pure states are made available whereas the
second method is optimally efficient even for a single pair. We would like to
stress that, although in principle, using the first qubit assisted method, one
can extract the optimal fraction of MES from a \textit{}finite ensemble of pure
states provided the iterative procedure is carried on indefinitely but this
iterative procedure makes sense in practice, only when Alice and Bob shares
a reasonably sized ensemble of pure states. We also suggested an entanglement
assisted concentration scheme which is also optimally efficient for a single
pair. We also discussed why a qubit assisted method is better than the entanglement
assisted one. Finally we have shown how these methods can be successfully used
to concentrate entanglement from multipartite cat like states. The concentration
probability thus obtained for N-party cat like states is found to be the same
as that in bipartite systems and is conjectured to be optimal. \\

\textbf{Acknowledgments}

I wish to thank Ujjwal Sen and Debasis Sarkar for useful comments and careful
reading of the manuscript. I am also thankful to an anonymous referee for pointing
out the experimental feasibility of different protocols.


\begin{thebibliography}{}
\bibitem{1}E. Schroedinger, Naturwissenschaften \textbf{23}, 807 (1935); \textbf{23}, 823
(1935); \textbf{23}, 844 (1935); For a review see: M. B. Plenio and V. Vedral,
Cont. Phys. \textbf{39}, 431 (1998); Lecture notes of Lucien hardy available
at http://www.qubit.org.
\bibitem{2}A. Einstein, B. Podolski, and N. Rosen, Phys. Rev. \textbf{47}, 777 (1935) {[}reprinted
in \textit{Quantum theory and Mesaurement}, edited by J. A. Wheeler and W. Z.
Zurek (Princeton University Press, Princeton, 1983){]}.
\bibitem{3}J. S. Bell, Physics \textbf{1}, 195 (1964) {[}reprinted in J. S. Bell, \textit{Speakable
and Unspeakable in Quantum Mechanics} (Cambridge University Press, Cambridge,
1987), p. 14{]}.
\bibitem{4}C. H. Bennett, Physics Today \textbf{48}, 24 (1995); J. Preskill, Proc. Roy.
Soc. A: Math., Phys. and Eng. \textbf{454}, 469 (1998).
\bibitem{5}A brief but excellent article is by R. Jozsa, quant-ph/9707034
\bibitem{6}C. H. Bennett, G. Brassard, C. Crepeau, R. Jozsa, A. Peres and W. K. Wootters,
Phys. Rev. Lett. \textbf{70}, 1895 (1993).
\bibitem{7}C. H. Bennett and S. J. Wiesner, Phys. Rev. Lett. \textbf{69}, 2882 (1992).
\bibitem{8}A. K. Ekert, Phys. Rev. Lett. \textbf{67}, 661 (1991).
\bibitem{9}H. Barnum, R. Cleve, and W. van Dam, quant-ph/9705033.
\bibitem{10}C. H. Bennett, H. J. Bernstein, S. Popescu, and B. Schumacher, Phys. Rev. A
\textbf{53}, 2046 (1996).
\bibitem{11}M. Murao, M. B. Plenio, S. Popescu, V. Vedral and P. L. Knight, Phys. Rev. A
\textbf{57}, 4075 (1998). 
\bibitem{12}S. Bose, V. Vedral and P. L Knight, Phys. Rev. A \textbf{60}, 194 (1999). Also
at LANL archive, quant-ph/9812013.
\bibitem{13}H. K. Lo and S. Popescu, quant-ph/9707038.
\bibitem{14}C. H. Bennett, D. P. DiVincenzo, J. A. Smolin and W. K. Wootters, Phys. Rev.
A \textbf{54}, 3824 (1996).
\bibitem{15}C. H. Bennett, G. Brassard, S. Popescu, B. Schumacher, J. A. Smolin and W. K.
Wootters, Phys. Rev. Lett. 76, 722 (1996); D. Deutsch, A. Ekert, R. Jozsa, C.
Macchiavello, S. Popescu and A. Sanpera, Phys. Rev. Lett. 77, 2818 (1996).
\bibitem{16}M. Zukowski, A. Zeilinger, M. A. Horne, and A. K. Ekert, Phys. Rev. Lett. \textbf{71},
4287 (1993); For multiparticle generalizations see, S. Bose, V. Vedral and P.
L. Knight, Phys. Rev. A \textbf{57}, 822 (1998).
\bibitem{17}K. Kraus, States, Effects and Operations: Fundamental Notions of Quantum theory
(Springer-Verlag, Berilin/New York, 1983); A. Peres, Quantum Theory: Concepts
and Methods (Kluwer, Dordrecht, 1993), Ch. 9.
\end{thebibliography}
\end{document}